\documentclass[preprint,showpacs,preprintnumbers,amsmath,amssymb]{revtex4}

\usepackage{graphicx}

\begin{document}

\title{Exact numerical simulation of power-law noises}
\author{Edoardo Milotti}
\email{milotti@ts.infn.it}
\affiliation{Dipartimento di Fisica, Universit\`a di Udine and I.N.F.N. -- Sezione di Trieste\\
Via delle Scienze, 208 -- I-33100 Udine, Italy}

\date{\today}

\begin{abstract}
Many simulations of stochastic processes require colored noises: I describe here an exact numerical method to simulate power-law noises: the method can be extended to more general colored noises, and  is exact for all time steps, even when they are unevenly spaced (as may often happen for astronomical data, see e.g. N. R. Lomb, Astrophys. Space Sci. {\bf 39}, 447 (1976)). The algorithm has a well-behaved computational complexity, it produces a nearly perfect Gaussian noise, and its computational efficiency depends on the required degree of noise Gaussianity. 
\end{abstract}

\maketitle


In recent years colored noise sources have been considered in many disparate applications, that range from stochastic resonance \cite{ghjm}, to biophysics \cite{lind,wenn} and beam dynamics in particle accelerators \cite{BS,SB}. The analytical approach to some of these processes is often difficult, and sometimes impossible, and numerical experiments are commonly used to support the analytical conclusions, or as an aid to discover new results. For this reason, algorithms that produce colored noise have acquired an ever increasing importance. 
This widespread interest spans different scientific communities, and the existing algorithms reflect the variety of approaches to the understanding of stochastic processes in different contexts. There are physics-inspired algorithms that rely mostly on equations of the Langevin type, FFT-based and autocorrelation function methods that use the spectral or correlation properties of colored noise, and time-series methods that produce colored noise from different filtering approaches. The review paper by Kasdin \cite{kas} provides a long list of references until 1995, centered mostly on linear processes and FFT methods. More recently, Greenhall wrote a review paper on FFT-based methods \cite{green}, and reference \cite{TK} is another very clear paper on the same topic. 
I describe here an exact numerical simulation of power-law noises that can be extended to more general colored noises, and which is based on the classical argument proposed long ago by Bernamont to model $1/f^\alpha$ noise as a superposition of Ornstein-Uhlenbeck processes \cite{bernamont}. The synthesis of colored noise from a point process is clearly not new, because this kind of modeling dates as far back as 1909,  to the work of Campbell \cite{camp} (see also the famous paper by Rice \cite{rice});  more recently Teich, Lowen and collaborators have carried out extensive studies on point processes with long-tail pulse response functions \cite{LT,LT2}, and others have studied the synthesis of power-law spectra from nonlinear processes (see, e.g.,  \cite{K} for a model based on a multiplicative point process). The simulation methods described in \cite{kas,green} assume evenly distributed sampling steps, and the extension to uneven sampling is not trivial: however noneven sampling has many important applications (see, e.g. the classic papers by Lomb and Scargle \cite{LS,scar} on period analysis for irregularly sampled astronomical data, and two more recent references \cite{HD,tao}), and Gillespie discussed a method valid for the Ornstein-Uhlenbeck process and based on the Langevin equation in 1996 \cite{gill}. The algorithm proposed here is very general (it is not limited to the OU process), it is very easy to implement, it is valid for all time steps,  it has a well-behaved computational complexity, and produces a nearly perfect Gaussian noise, and its computational efficiency depends on the required degree of noise Gaussianity.

Here we take a signal $x(t)$ that originates from the linear superposition of many random pulses, i.e., pulses that are random in time and can be described by a memoryless process with a Poisson distribution, have random amplitude $A$ drawn from a distribution with finite variance and probability density  $g_A(A)$, and such that their pulse response function is 
\begin{equation}
\label{prf}
h(t,\lambda) = 
\left\{
\begin{array}{cc}
\exp(-\lambda t)  & \mathrm{ if } \; t \ge 0\\
0  &  \mathrm{ if } \; t<0
\end{array}
\right.
\end{equation}
with a decay rate which is drawn from a distribution with probability density $g_\lambda(\lambda)$, so that
\begin{equation}
\label{sum}
x(t) = \sum_k A_k h(t-t_k, \lambda_k)
\end{equation}
where $t_k$ is the time at which the $k$-th pulse occurs, $A_k$ is its amplitude and $\lambda_k$ is the decay rate of its pulse response function.

If $n$ is the pulse rate, then on average there are $n \left[g_A(A) dA\right]  \left[g_\lambda(\lambda) d\lambda\right]  dt$ pulses in the time interval $(t',t'+dt)$ and in the amplitude-$\lambda$ range $dA d\lambda$; the number of pulses follows a Poisson distribution and therefore the variance of the number of detected pulses is also equal to $n \left[g_A(A) dA\right]  \left[g_\lambda(\lambda) d\lambda\right]  dt$. This means that the mean and the variance of the output signal at time $t$ are given by the integrals
\begin{equation}
\label{mean}
\langle x \rangle = \int_{\lambda_{min}}^{\lambda_{max}} g_\lambda(\lambda) d\lambda \int_{A_{min}}^{A_{max}} g_A(A) dA  \int_{-\infty}^t dt' n \left[A h(t-t',\lambda) \right]
\end{equation}
and
\begin{equation}
\label{msq}
\langle (\Delta x)^2 \rangle  = \int_{\lambda_{min}}^{\lambda_{max}} g_\lambda(\lambda) d\lambda \int_{A_{min}}^{A_{max}} g_A(A) dA  \int_{-\infty}^t dt' n \left[A h(t-t',\lambda) \right]^2
\end{equation}
If we assume that the amplitude $A$ is fixed, we take the pulse response function (\ref{prf}), and rearrange the time integration, the integrals (\ref{mean}) and (\ref{msq}) simplify to 
\begin{equation}
\label{mean2}
\langle x \rangle  = n A  \int_{\lambda_{min}}^{\lambda_{max}} g_\lambda(\lambda) d\lambda \int_0^\infty dt  \left[ h(t,\lambda) \right] =  n A  \int_{\lambda_{min}}^{\lambda_{max}} \frac{g_\lambda(\lambda)}{\lambda} d\lambda = n A  \left\langle \frac{1}{\lambda} \right\rangle
\end{equation}
and
\begin{equation}
\label{msq2}
\langle (\Delta x)^2 \rangle  = n A^2  \int_{\lambda_{min}}^{\lambda_{max}} g_\lambda(\lambda) d\lambda \int_0^\infty dt  \left[ h(t,\lambda) \right] ^2=  n A^2  \int_{\lambda_{min}}^{\lambda_{max}} \frac{g_\lambda(\lambda)}{2\lambda} d\lambda = \frac{n A^2}{2}  \left\langle \frac{1}{\lambda} \right\rangle
\end{equation}
Now let $H(\omega,\lambda)$ be the Fourier transform of $h(t,\lambda)$, then from the causality constraint on $h(t,\lambda)$ and Parseval's theorem we find that the variance (\ref{msq2}) can be trasformed into 
\begin{equation}
\label{msq3}
\langle (\Delta x)^2 \rangle  = \frac{n A^2 }{2\pi} \int_{\lambda_{min}}^{\lambda_{max}} g_\lambda(\lambda) d\lambda \int_{-\infty}^\infty d\omega  \left| H(\omega,\lambda) \right|^2=\frac{n A^2}{2\pi} \int_{-\infty}^\infty d\omega  \int_{\lambda_{min}}^{\lambda_{max}} g_\lambda(\lambda) d\lambda \left| H(\omega,\lambda) \right|^2
\end{equation}
The right-hand expression in equation (\ref{msq3}) shows that the spectral density is 
\begin{equation}
\label{psd}
S(\omega) = \frac{n A^2}{2\pi} \int_{\lambda_{min}}^{\lambda_{max}} g_\lambda(\lambda) d\lambda \left| H(\omega,\lambda) \right|^2
\end{equation}
and since $\left| H(\omega,\lambda) \right|^2=(\omega^2 + \lambda^2)^{-1}$ for the exponential pulse response function (\ref{prf}), we obtain eventually 
\begin{equation}
\label{psd2}
S(\omega) = \frac{n A^2}{2\pi} \int_{\lambda_{min}}^{\lambda_{max}} \frac{g_\lambda(\lambda)}{\omega^2 + \lambda^2} d\lambda
\end{equation}

We consider now three special, important cases: if there is just a single decay rate $\lambda$, the  spectral density has the usual Lorentzian shape
\begin{equation}
\label{debye}
S(\omega) = \frac{n A^2}{2\pi}  \frac{1}{\omega^2 + \lambda^2} 
\end{equation}
which, for $\omega \gg \lambda$ has a  $1/f^2$ behavior, so that we can approximate a $1/f^2$ spectrum in an actual process by choosing a $\lambda$ smaller than the lowest observed frequency. 
With a careful choice of the distribution $g_\lambda(\lambda)$ we can synthesize many different spectra, but there are two special choices for $g_\lambda(\lambda)$: we can take a uniform distribution or a range-limited power-law distribution. If we assume a uniform distribution of decay rates, between $\lambda_{min}$ and $\lambda_{max}$, i.e. 
\begin{equation}
\label{unif}
g_\lambda(\lambda) = \frac{1}{\lambda_{max} - \lambda_{min}}
\end{equation}
then the average $\langle 1/\lambda \rangle$ that determines the mean level (\ref{mean2}), and the variance (\ref{msq2}) is 
\begin{equation}
\left\langle \frac{1}{\lambda} \right\rangle = \frac{\ln(\lambda_{max}/\lambda_{min})}{\lambda_{max}-\lambda_{min}}
\end{equation}
and using equation \ref{psd2}, the spectral density is easily shown to be 
\begin{equation}
\label{psdflat}
S(\omega) = \frac{n A^2 }{2\pi(\lambda_{max}-\lambda_{min})} \frac{1}{\omega} \left(\arctan\frac{\lambda_{max}}{\omega}-\arctan\frac{\lambda_{min}}{\omega}\right)
\end{equation}
and in the range $\lambda_{min} \ll \omega \ll \lambda_{max}$ this spectral density has a $1/f$ behavior. Similarly, if we take a range-limited power-law distribution
\begin{equation}
\label{plaw}
g_\lambda(\lambda) = \left(\frac{1-\beta}{\lambda_{max}^{1-\beta} - \lambda_{min}^{1-\beta}}\right) \lambda^{-\beta}
\end{equation}
then the average $\langle 1/\lambda \rangle$ is 
\begin{equation}
\left\langle \frac{1}{\lambda} \right\rangle = -\left(\frac{1-\beta}{\beta}\right)\frac{\lambda_{max}^{-\beta}-\lambda_{min}^{-\beta}}{\lambda_{max}^{1-\beta}-\lambda_{min}^{1-\beta}}
\end{equation}
and the spectral density is
\begin{eqnarray}
\label{psdplaw}
\nonumber
S(\omega) &=& \frac{1}{(\lambda_{max}^{1-\beta} - \lambda_{min}^{1-\beta})\omega^2}\left[  \lambda_{max}^{1-\beta} F\left(\frac{1-\beta}{2},1;\frac{1-\beta}{2};\frac{-\lambda_{max}^2}{\omega^2}\right) \right. \\ && \left.- \lambda_{min}^{1-\beta} F\left(\frac{1-\beta}{2},1;\frac{1-\beta}{2};\frac{-\lambda_{min}^2}{\omega^2}\right)  \right]
\end{eqnarray}
which has a $1/f^{1+\beta}$ behavior in the range $\lambda_{min} \ll \omega \ll \lambda_{max}$.

\begin{figure}
\includegraphics[width=5in]{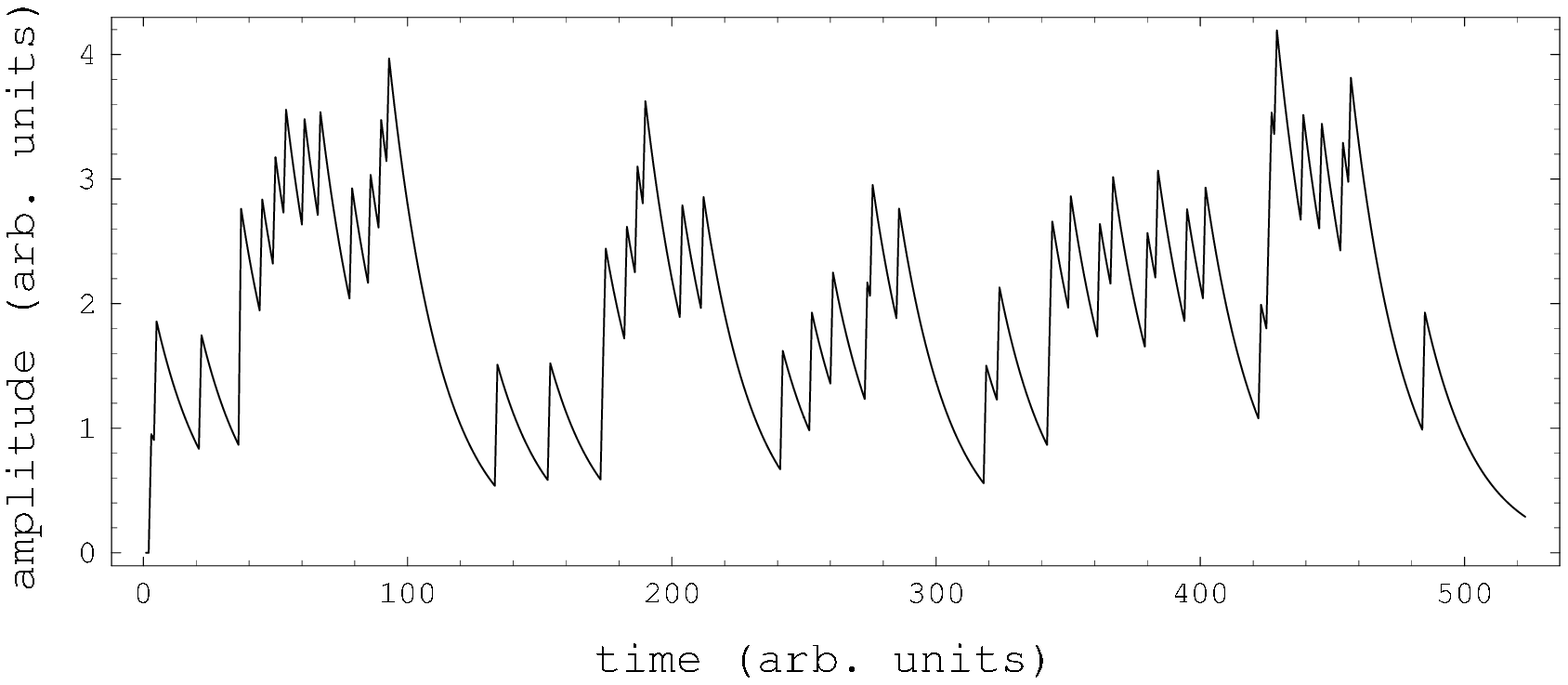}
\caption{\label{fig1} A realization of the random process $x(t)$ (eq. \ref{sum}) with $A=1$, $n=1$, and with fixed decay rate $\lambda = 0.5$ (all quantities are given in arbitrary units; $\lambda$ is given in inverse time units). The single exponential decays are clearly visible, and the random process is obviously non-Gaussian.
}
\end{figure}

Now we follow the lead provided by these considerations,  and we take, e.g., the case where there is just a single decay rate $\lambda$, so that the spectral density has the Lorentzian shape (\ref{debye}). Since the probability density of the time intervals $\Delta t_k$ between Poisson events is well-known to be $dP(\Delta t) = n \exp(-n\Delta t) d\Delta t$, we can generate a sequence of $\Delta t$'s from an exponential distribution, and we can thus generate the sequence $\{t_k\}$ (with $t_{k+1} = t_k + \Delta t_k$) required to evaluate a realization of $x(t)$ as in equation (\ref{sum}): figure \ref{fig1} shows an example where the single decays are clearly visible. Figure \ref{fig1} also shows that, although the process has the desired spectral density, it is quite obviously non-Gaussian and therefore this generation method seems to be of marginal utility, as most of the actual physical processes are Gaussian and Gaussianity is usually a required property of a good noise generator (see, e.g., the recent paper \cite{lee} that describes a hardware-based Gaussian white noise simulator and contains a list of relevant references; notice also that Gaussianity is sometimes a weakness rather than a strength, see, e.g. \cite{KL}). The Gaussianity in shot noise processes has been studied at length since the paper by Rice \cite{rice} and here we strictly limit the discussion to the special processes considered in this paper. The single exponential spikes in figure \ref{fig1} stand out more clearly when the average rate $n$ of the Poisson process is smaller than the decay rate $\lambda$; by contrast, when $n \gg  \lambda$, at any time there are many pulses of comparable size and the sum has a nearly Gaussian behavior. We can gain further insight in this generation method by using the mean moment generating function (mmgf) for a Poisson process with average rate $a$: 
\begin{eqnarray}
\label{mmgf }
\langle \exp\left[ i t (k-\langle k \rangle) \right] \rangle & = & \sum_{m=0}^{\infty} \langle (k-\langle k \rangle )^m \rangle \frac{(i t)^m}{m!} =
\exp \left[ (e^{it}-1)a - ita \right] \\
\nonumber
& = & 1 + \frac{i^2}{2!} at^2 + \frac{i^3}{3!} at^3 + \frac{i^4}{4!} (3a^2+a) t^4 \\
& & + \frac{i^5}{5!} (10a^2+a) t^5 + \frac{i^6}{6!} (15a^3 + 5a^2 + a) t^6 + O(t^7)
\end{eqnarray}
Now we use the mmgf to compute the higher-order moments: as already discussed in the derivation of equations (\ref{mean}) and (\ref{msq}), the process $x(t)$ is the sum of Poisson variates with different amplitudes; on the other hand the mmgf of the weighted sum $\alpha k + \beta j$  of two independent Poisson variates $k$ and $j$, both with rate $a$, is 
\begin{eqnarray}
\langle \exp \left\{i t \left[ (\alpha k + \beta j)-\langle \alpha k + \beta j \rangle  \right] \right\} \rangle & = & \langle \exp \left[i t \alpha (k - \langle k  \rangle  \right]  \rangle \langle \exp \left[i t \beta (j - \langle j  \rangle  \right]  \rangle \\
\nonumber
& = & 1 +  \frac{i^2}{2!}a \left[ (\alpha t)^2 + (\beta t)^2 \right] \\
\label{mmgfP}
& & + \frac{i^3}{3!} \left[ (\alpha t)^3 + (\beta t)^3 \right] + O(t^4)
\end{eqnarray}
Using the mmgf's given above we could proceed as in standard texts on probability theory, and show that for large $a$ the process approaches an exact Gaussian distribution (the usual proof of the Central Limit Theorem), but the purpose here is giving a quantitative estimate of the {\em deviation} from Gaussianity: from the expansion (\ref{mmgfP}), we see that, just like the variance, the third moment about the mean of the weighted sum of independent Poisson variates is the weighted sum of the third moments of the individual variates (this is not true for the fourth and the higher moments), and therefore we can write a simple expression for the third moment about the mean
\begin{equation}
\label{third}
\langle (\Delta x)^3 \rangle  = n A^3  \int_{\lambda_{min}}^{\lambda_{max}} g_\lambda(\lambda) d\lambda \int_0^\infty dt  \left[ h(t,\lambda) \right]^3
\end{equation}
and we can use this expression to compute the skewness of the frequency distribution 
\begin{equation}
\label{skew}
\mathrm{skewness} = \frac{\langle (\Delta x)^3 \rangle}{\langle (\Delta x)^2 \rangle^{3/2}} = \frac{\int_{\lambda_{min}}^{\lambda_{max}} g_\lambda(\lambda) d\lambda \int_0^\infty dt  \left[ h(t,\lambda) \right]^3}{ \sqrt{n} \left\{ \int_{\lambda_{min}}^{\lambda_{max}} g_\lambda(\lambda) d\lambda \int_0^\infty dt  \left[ h(t,\lambda) \right]^2 \right\}^{3/2}}
\end{equation}
With the pulse response function (\ref{prf}) the integrals are easily evaluated, so that 
\begin{equation}
\label{skew2}
\mathrm{skewness} = \frac{\int_{\lambda_{min}}^{\lambda_{max}} \left[ g_\lambda(\lambda)/(3\lambda)\right] d\lambda }{ \sqrt{n} \left\{ \int_{\lambda_{min}}^{\lambda_{max}} \left[ g_\lambda(\lambda)/(2\lambda\right] d\lambda \right\}^{3/2}} = \frac{2^{3/2}}{3} \frac{1}{\sqrt{n \langle 1/\lambda \rangle}}
\end{equation}
From equations (\ref{skew}) and (\ref{skew2}), we see that the skewness is small when $n\langle 1/\lambda \rangle$ is large (as it should be for a Gaussian distribution) and that the actual amount of skewness depends on the adimensional product $n \langle 1/\lambda \rangle$, as expected; as a rule of thumb one might take $n \langle 1/\lambda \rangle > 10$ for good Gaussianity.

\begin{figure}
\includegraphics[width=5in]{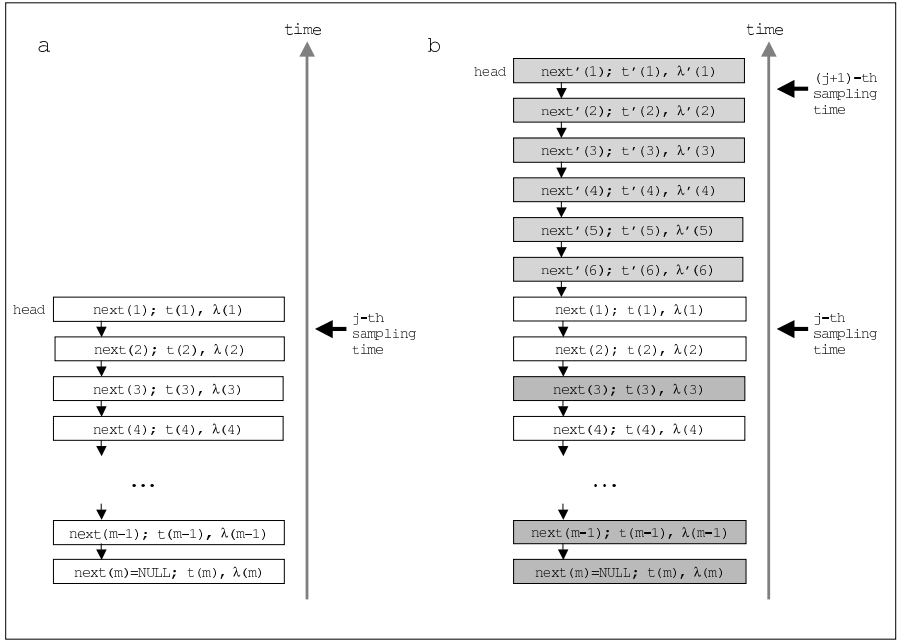}
\caption{\label{fig2} Structure and update dynamics of the linked list that holds the Poisson-distributed transition events: {\bf a}. structure of the list at the $j$-th sampling time $s_j$; each node contains a variable that points to the next node (the end of the list is marked by a null pointer), and stores the time $t_k$  of the transition and the decay rate $\lambda_k$ of the associated pulse response function. The list contains only nodes such that $s_j-t_k \le N_{decay}/\lambda_k$. The response of the system is computed from the sum  $\sum_k \exp[-\lambda_k (s_j-t_k)]$, where the index $k$ ranges over all the list elements such that $s_j-t_k \ge 0$ (the list head is usually excluded). {\bf b}. if the $(j+1)$-th sampling time is greater than the time stored in the list head (as is usually the case), the program generates as many transition times as needed to reach (and possibly overcome) the $(j+1)$-th sampling time (light-gray boxes in the figure, primed quantities), and next it scans the list to discard all the nodes such that $s_{j+1}-t_k > N_{decay}/\lambda_k$ (dark-gray boxes). At this point the program computes the new response and steps to the next sampling step.
}
\end{figure}

The previous considerations apply to the noise process $x(t)$ without any reference to sampling, however the simulation of noisy physical systems usually implies evaluating the noise process at evenly spaced sampling times, so we take now a sequence of sampling times $\{s_j\}$ with average sampling interval $\langle \Delta s \rangle$. 
At each sampling time only the recent pulses actually contribute, while the older pulses quickly fade away, for instance the average total contribution of pulses that are older than $N_{decay}/\lambda$ is 
\begin{eqnarray}
\label{oldmean}
\nonumber
\langle \delta x \rangle &=& n A  \int_{\lambda_{min}}^{\lambda_{max}} g_\lambda(\lambda) d\lambda\int_{-\infty}^{t-N_{decay}/\lambda} dt' h(t-t',\lambda)\\
 & = & n A \int_{\lambda_{min}}^{\lambda_{max}} g_\lambda(\lambda) d\lambda \int_{N_{decay}/\lambda}^\infty h(t'')dt''  = n A \left \langle \frac{1}{\lambda} \right \rangle e^{-N_{decay}}
\end{eqnarray}
which is just a small fraction of the mean value: $\langle \delta x \rangle / \langle x \rangle = e^{-N_{decay}}$, and for this reason as we proceed forward in time, we can just forget the older transitions. In an actual implementation we fix $N_{decay}$, but because of random event clustering we cannot know {\it a priori} how many transitions times must actually be kept in memory: for this reason the Poisson-distributed transition times should not be stored in an array, but in a linked list \cite{NS}; the linked list must also store the decay rates that correspond to each transition event. At each sampling step the list is updated first by generating as many transition times (and the associated decay rates, which are drawn from a given decay rate distribution) as needed to reach (and possibly surpass) the actual sampling time $s_j$, and then by discarding those events with an occurrence time $t_k$ such that $s_j - t_k > N_{decay}/\lambda$ (see figure \ref{fig2}). The mean list length is just $n N_{decay}\langle 1/\lambda \rangle$, and the processing time is proportional to the number of list elements. At startup the list is empty, and the first $N_{decay}/\lambda_{min} \langle \Delta s \rangle$ samples must be used for initialization and afterwards discarded as the algorithm fills the list up to the average level, and thus, for a desired number of samples $N_s$, we must generate a total of $N_s + N_{decay}/(\lambda_{min} \langle \Delta s \rangle)$ samples.
The time-complexity of the algorithm is thus proportional to the sum of the total number of generated transitions plus the total number of operations used for the list scans, i.e.,
\begin{eqnarray}
\nonumber
\mathrm{complexity} & = & \left( N_s + \frac{N_{decay}}{\langle \Delta s \rangle} \frac{1}{\lambda_{min}}  \right) \left( C_1 n \langle \Delta s \rangle + C_2 n N_{decay}\left\langle \frac{1}{\lambda} \right \rangle \right)\\
&  = & n \langle \Delta s \rangle \left( N_s + \frac{N_{decay}}{\langle \Delta s \rangle} \frac{1}{\lambda_{min}} \right) \left( C_1  + C_2  \frac{N_{decay}}{\langle \Delta s \rangle}\left\langle \frac{1}{\lambda} \right \rangle \right) 
\end{eqnarray}

\begin{figure}
\includegraphics[width=5in]{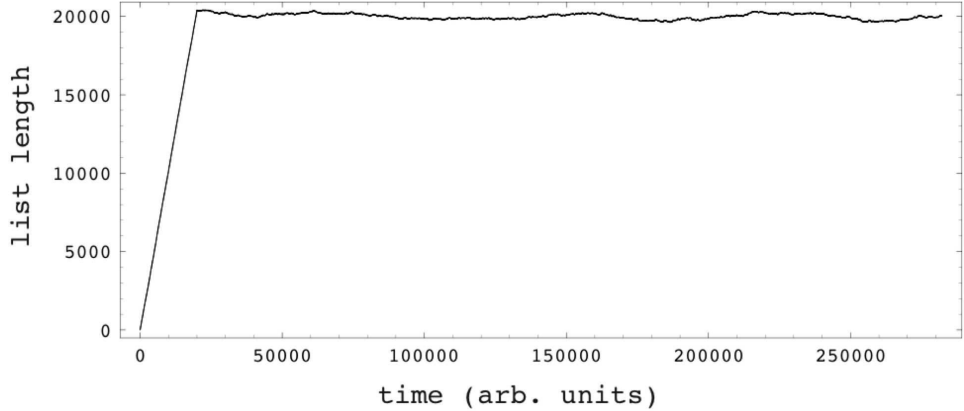}
\caption{\label{fig3} Length of the linked list in a simulation with $A=1$ and a single decay rate $\lambda = 0.001$: the linked list is initially empty, at it fills up at a constant rate. In this case $n=1$, $N_{decay} = 20$ and $\Delta s = 1$ and therefore the fill-up time is $ (N_{decay}/\lambda) = 20000$ the fill-up length is $n (N_{decay}/\lambda) = 20000$, and the number of samples required for the initial fill-up  is $ (N_{decay}/\lambda)/\Delta s = 20000$. After the initial fill-up the length of the linked list fluctuates about the average filling level.
}
\end{figure}

\begin{figure}
\includegraphics[width=5in]{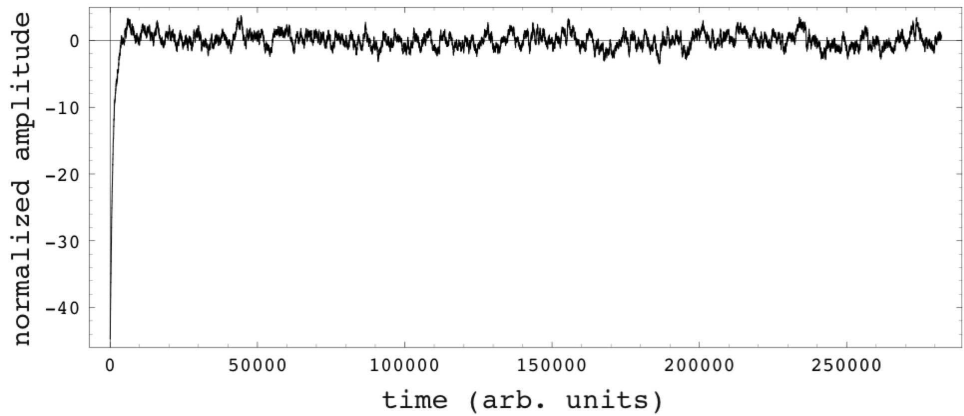}
\caption{\label{fig4} Plot of the normalized signal amplitude $(x(t)-\langle x \rangle)/\sigma$ ($\sigma$ is the standard deviation of the amplitude, i.e., the square root of the variance (\ref{msq2}) ) in the simulation run described in figure \ref{fig3} and in the text. At the beginning the linked list which contains the process memory is empty, and the signal is very far off the predicted average; as the list fills up to level, the signal quickly reaches the predicted average.
}
\end{figure}

\begin{figure}
\includegraphics[width=5in]{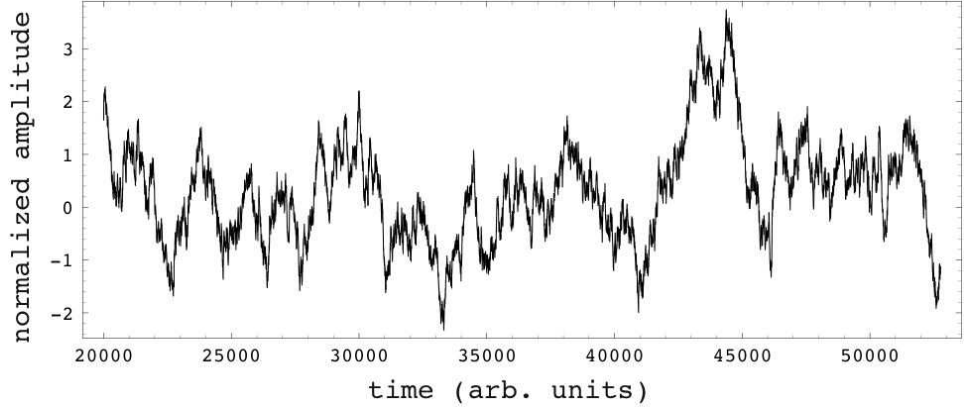}
\caption{\label{fig5} Detail of the normalized signal amplitude shown in the figure \ref{fig4}, just after the list has filled up to the average level. This signal displays the large characteristic upward an downward swings that are well-known in the theory of random walks \cite{feller}.
}
\end{figure}

\begin{figure}
\includegraphics[width=5in]{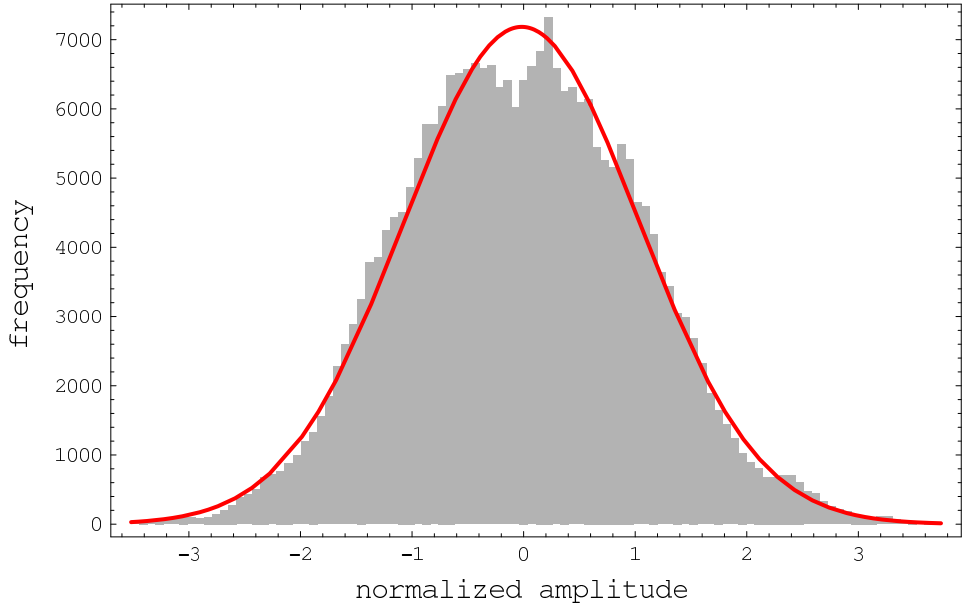}
\caption{\label{fig6} The histogram shows the amplitude distribution of 262144 samples from the realization of the random process $x(t)$ shown in figure \ref{fig4}, after the list fill-up. The continuous curve is a Gaussian with the mean and standard deviation estimated from the samples. There is no visible skewness, because in this simulation run $\lambda/n = 0.001$, which corresponds to a very low skewness (\ref{skew2}), but there are multiple peaks, which are due to the nonstationarity of a true $1/f^2$ process (which is well approximated here), and which require an extremely long observation time to establish the Gaussianity of the process \cite{feller}. 
}
\end{figure}

\begin{figure}
\includegraphics[width=5in]{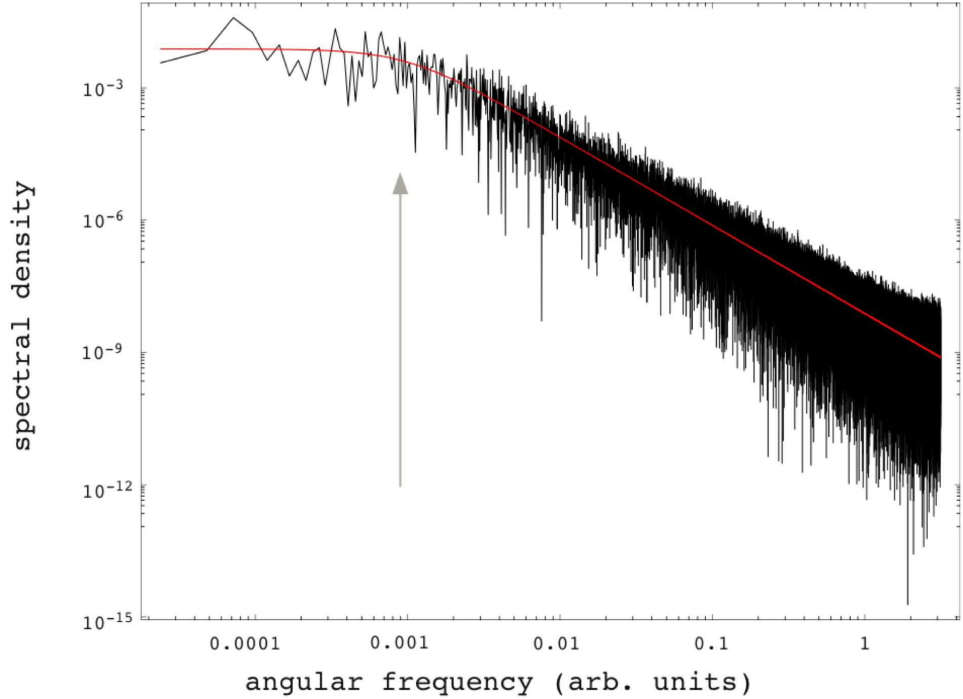}
\caption{\label{fig7} DFT spectrum obtained from 262144 samples from the realization of the normalized signal amplitude $(x(t)-\langle x \rangle)/\sigma$ shown in figure \ref{fig4}. The continuous curve shows the theoretical power spectral density (\ref{debye}). Because of sampling without low-pass filtering there is some aliasing and the DFT spectrum shows a slight upward bend at high frequency. Since the sampling interval is $\Delta s = 1$, the Nyquist (angular) frequency is just $\omega_{Nyquist} = \pi$ (here and in the following spectra time is measured in arbitrary units as in the previous figures, and frequency units are defined accordingly). The arrow marks the position of the single decay rate in this simulation $\lambda = 0.001$. 
}
\end{figure}

\begin{figure}
\includegraphics[width=5in]{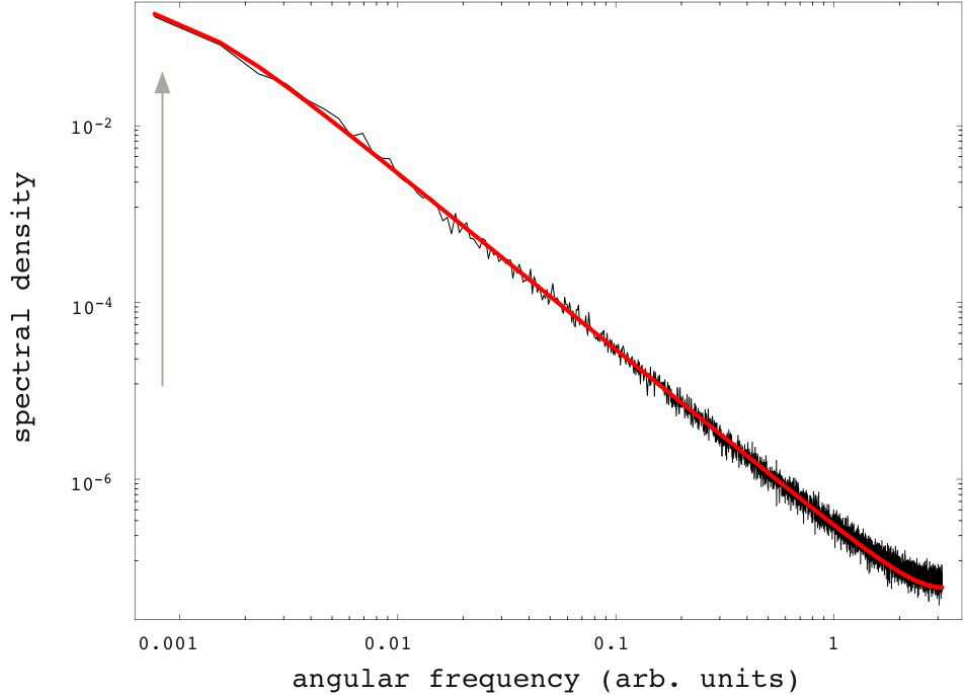}
\caption{\label{fig8} Averaged DFT spectrum obtained from the same 262144 samples as the spectrum in figure \ref{fig7}, split in 32 blocks of 8192 samples each. The continuous curve shows the theoretical power spectral density (\ref{debye}), and now it includes also the first-order correction to aliasing. Because of the low-frequency correlation between the blocks (that have been obtained from the same simulation record), the average spectrum is a bit higher than expected and the theoretical prediction has been globally shifted 20\% higher to fit the average spectrum; this artifact is absent in the analysis of the whole record (the low-frequency plateau of the spectrum in figure \ref{fig7} fits the theoretical curve exactly as expected). As in figure \ref{fig7}, the arrow marks the position of the single decay rate in this simulation $\lambda = 0.001$: because of the shorter record length used for DFT analysis, the frequency resolution is poorer here, and the spectrum mimics quite well the behavior of a true $1/f^2$ spectrum, over about three frequency decades.
}
\end{figure}

The algorithm described above is easily implemented;  figures \ref{fig3} to \ref{fig8} show the results obtained in a simulation of $N_s = 2^{18} = 262144$ transitions, with a single decay rate $\lambda = 0.001$, and a Poisson transition rate $n=1$ (here and in all the following discussions the system of units is arbitrary); moreover $N_{decay} = 20$, so that the average relative error due to the past transitions that have been discarded is $\langle \delta x \rangle/\langle x \rangle = \exp(-N_{decay}) \approx 2\cdot 10^{-9}$. With these parameters we expect an average list length $n N_{decay} /\lambda = 20000$, and a corresponding filling time $N_{decay} /(\lambda \Delta s) = 20000$: figure \ref{fig3} shows the list length, which behaves exactly as expected. Figures \ref{fig4} and \ref{fig5} show  the normalized signal amplitude $(x(t)-\langle x \rangle)/\sigma$, where $\sigma$ is the standard deviation of the amplitude, i.e., the square root of the variance (\ref{msq2}): at the beginning the linked list which contains the process memory is empty, and the signal is very far off the predicted average, but as the list fills up to level, the signal quickly reaches the predicted average. Figure \ref{fig6} is the histogram of the normalized signal amplitude obtained from 262144 samples, after the list fill-up; the continuous curve superimposed on the histogram is a Gaussian with the mean and standard deviation estimated from the samples, and we see that there is no visible skewness, because in this simulation run $\lambda/n = 0.001$, which corresponds to a very low skewness (\ref{skew2}), but there are multiple peaks, which are due to the nonstationarity of a true $1/f^2$ process (which is well approximated here), and which require an extremely long observation time to establish the Gaussianity of the process \cite{feller}. Finally figures \ref{fig7} and \ref{fig8} show the Discrete Fourier Transform (DFT) spectrum of the normalized signal amplitude, which reproduces quite well the expected shape (\ref{debye}).

\begin{figure}
\includegraphics[width=5in]{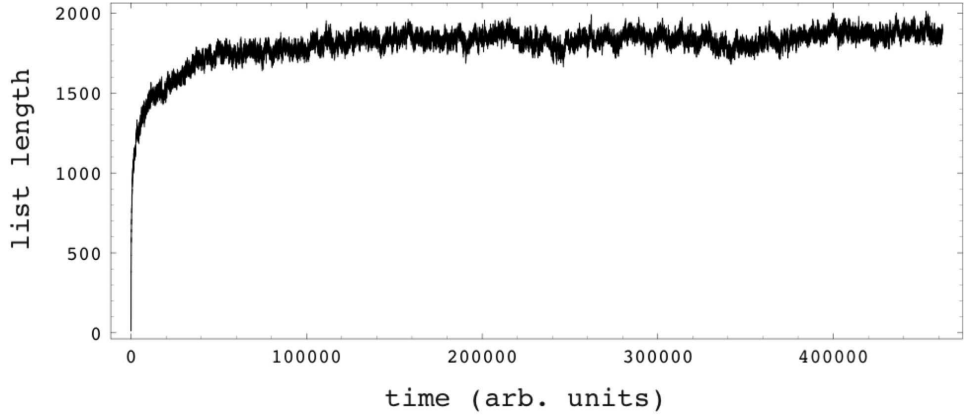}
\caption{\label{fig9} Length of the linked list in a simulation with $A=1$ and a uniform distribution of decay rates in the range  $\lambda_{min} = 0.0001$, $\lambda_{max} = 1$: the linked list is initially empty, at it fills up with a variable rate that depends on the distribution of decay rates. In this case $n=10$, $N_{decay} = 20$ and $\Delta s = 1$ and $\langle 1/\lambda \rangle \approx 9.211$, and therefore the fill-up time is $ N_{decay}/\lambda_{min} = 200000$ the fill-up length is $n (N_{decay}\langle 1/\lambda \rangle) \approx 1842.$, and the number of samples required for the initial fill-up  is $ (N_{decay}/\lambda_{min})/\Delta s = 200000$. After the initial fill-up the length of the linked list fluctuates about the average filling level.
}
\end{figure}

\begin{figure}
\includegraphics[width=5in]{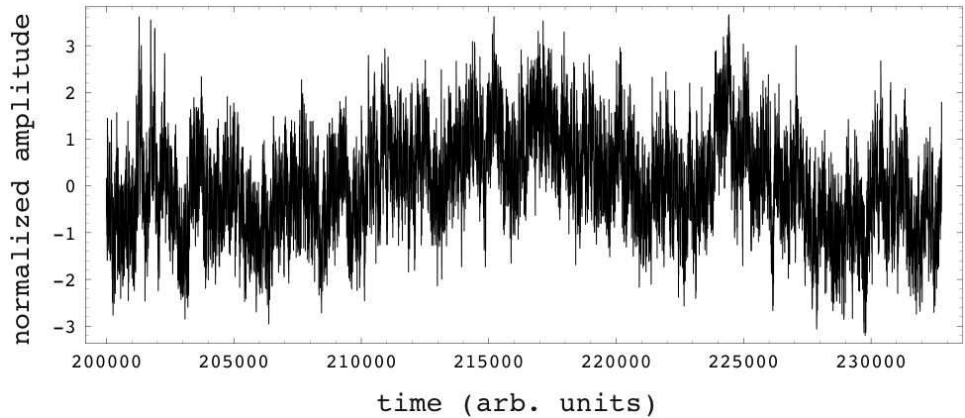}
\caption{\label{fig10} Detail of the normalized signal amplitude in the simulation of figure \ref{fig9}, just after the list has filled up to the average level.
}
\end{figure}

\begin{figure}
\includegraphics[width=5in]{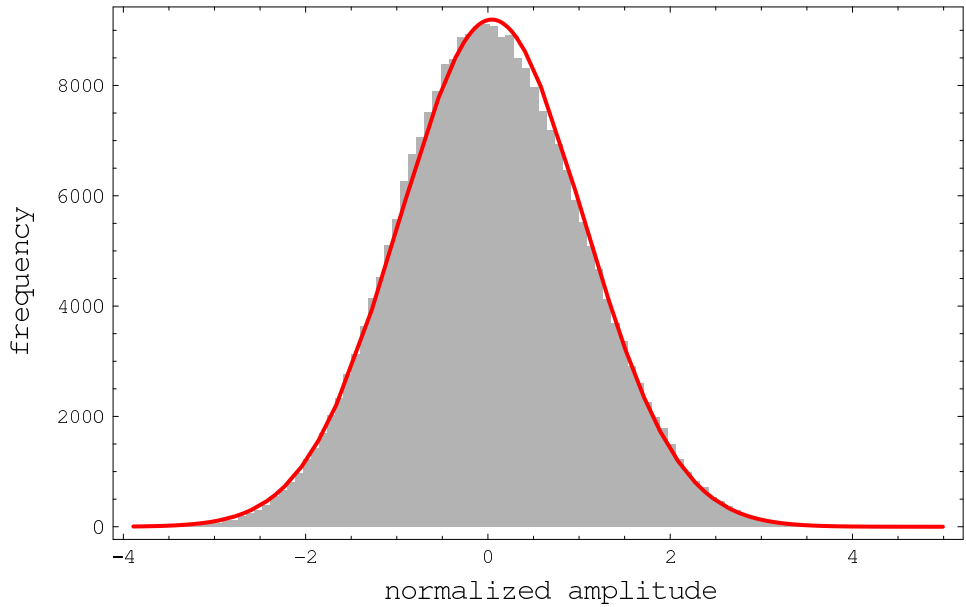}
\caption{\label{fig11} The histogram shows the amplitude distribution of 262144 samples from the realization of the random process $x(t)$ shown in figure \ref{fig10}, after the list fill-up. The continuous curve is a Gaussian with the mean and standard deviation estimated from the samples. In contrast to the histogram in figure \ref{fig6}, now the amplitude distribution appears slightly skewed, because in this simulation run $1/(n \langle 1/\lambda \rangle) \approx 0.0109$, noticeably higher than the corresponding value for figure \ref{fig6}. 
}
\end{figure}

\begin{figure}
\includegraphics[width=5in]{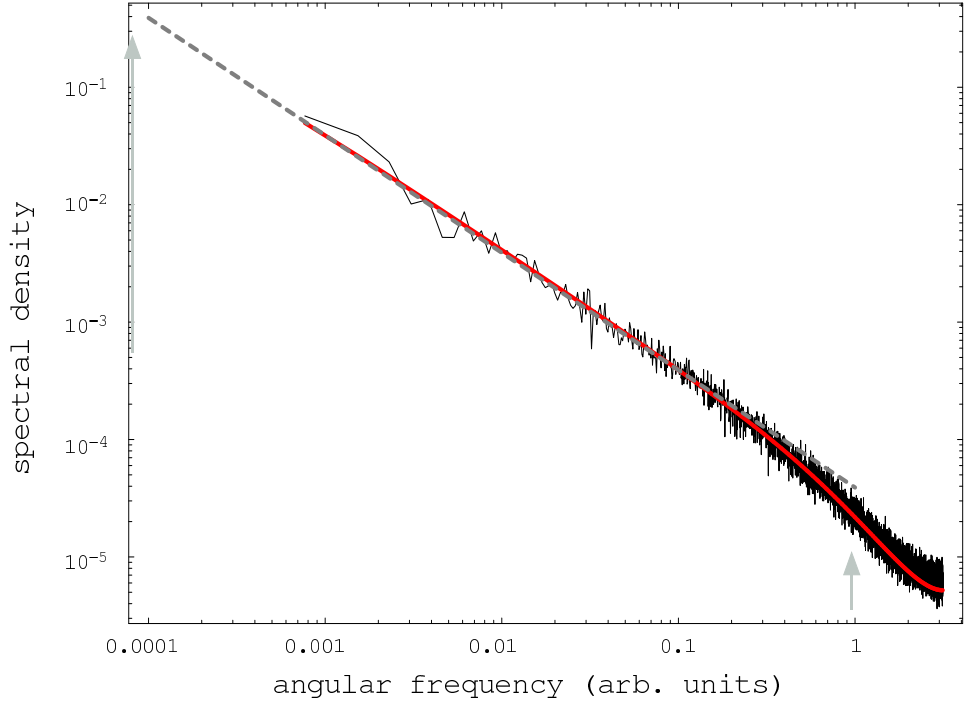}
\caption{\label{fig12} Averaged DFT spectrum obtained from the 262144 samples in the simulation of figure \ref{fig9}, split in 32 blocks of 8192 samples each. The continuous curve shows the theoretical power spectral density (\ref{psdflat}), which includes also the first-order correction to aliasing. The arrows mark the positions of the extreme decay rates $\lambda_{min} = 0.0001$ and $\lambda_{max} = 1$. The spectral resolution $\Delta \omega \approx 0.0015$ is larger than the minimum decay rate $\lambda_{min}$, and the spectrum mimics quite well the behavior of a true $1/f$ spectrum (dashed line), over more than three frequency decades.
}
\end{figure}

The next set of figures shows the results of a simulation with a decay rates that are uniformly distributed between $\lambda_{min} = 0.0001$ and $\lambda_{max} = 1$. Just as before, the simulation contains $N_s = 2^{18} = 262144$ samples, with a transition rate $n=10$, a sampling interval $\Delta s = 1$ and $N_{decay} = 20$: from these parameters we obtain the average $\langle 1/\lambda \rangle \approx 9.211$, so that the fill-up time is $ N_{decay}/\lambda_{min} = 200000$, the fill-up length is $n (N_{decay}\langle 1/\lambda \rangle) \approx 1842.$, and the number of samples required for the initial fill-up  is $ (N_{decay}/\lambda_{min})/\Delta s = 200000$. Figure \ref{fig9} shows the linked list length, which in this case does not reach the average filling level with a linear growth law, but with a smoothed curve. Figure \ref{fig10} shows the initial part of the simulated signal, and figure \ref{fig11} shows the histogram of the normalized signal amplitude:  in contrast to the histogram in figure \ref{fig6}, now the amplitude distribution is slightly skewed, because in this simulation run $1/(n \langle 1/\lambda \rangle) \approx 0.109$, much higher than the calculated skewness for figure \ref{fig6}.  Finally the averaged DFT spectrum is shown in figure \ref{fig12}: the spectrum mimics quite well the behavior of a true $1/f$ spectrum over more than three frequency decades.

\begin{figure}
\includegraphics[width=5in]{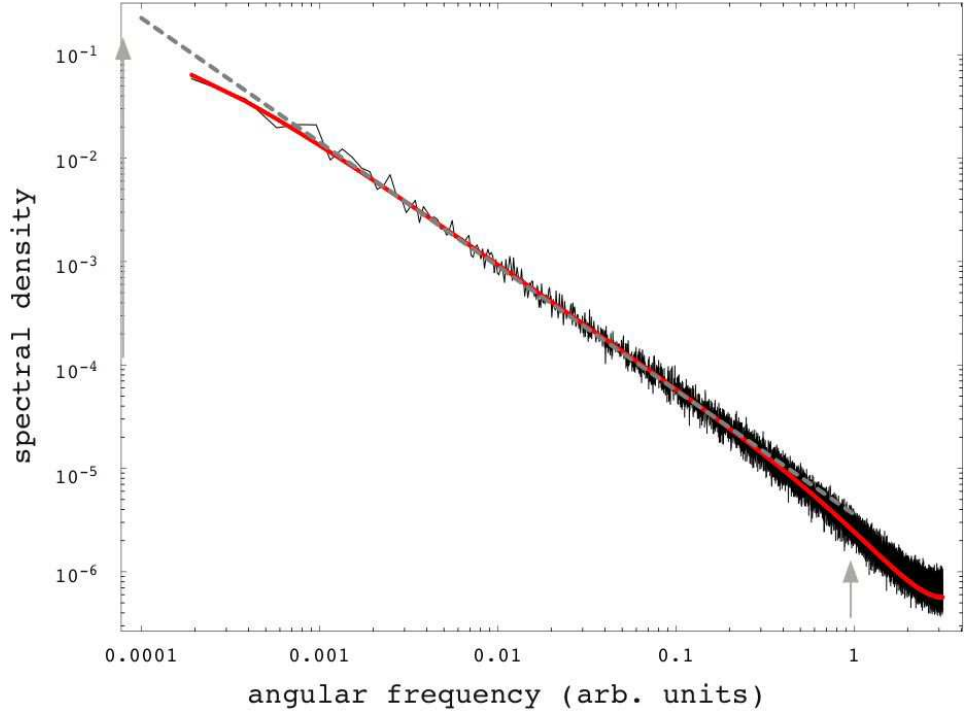}
\caption{\label{fig13} Averaged DFT spectrum obtained from $2^{20} = 1048576$ samples with $A=1$ and a range-limited power-law distribution of decay rates $\lambda^{-\beta}$, with $\beta = 0.2$, in the range  $\lambda_{min} = 0.0001$, $\lambda_{max} = 1$, split in 32 blocks of 32768 samples each. The continuous curve shows the theoretical power spectral density (\ref{psdplaw}), which includes also the first-order correction to aliasing. The arrows mark the positions of the extreme decay rates $\lambda_{min} = 0.0001$ and $\lambda_{max} = 1$. The spectral resolution $\Delta \omega \approx 0.00038$ is larger than the minimum decay rate $\lambda_{min}$, and the spectrum mimics quite well the behavior of a true $1/f^{1.2}$ spectrum (dashed line), over more than three frequency decades.
}
\end{figure}

The  \ref{fig13} shows the average spectrum obtained in a long simulation run with a different power-law noise:  $N_s = 2^{20} = 1048576$ samples have been generated with $A=1$ and a range-limited power-law distribution of decay rates (\ref{plaw}) with $\beta = 0.2$, in the range  $\lambda_{min} = 0.0001$, $\lambda_{max} = 1$. Here too the spectral resolution $\Delta \omega \approx 0.00038$ is larger than the minimum decay rate $\lambda_{min}$, and the noise samples reproduce the behavior of a true $1/f^{1.2}$ spectrum (dashed line), over more than three frequency decades.

\begin{figure}
\includegraphics[width=5in]{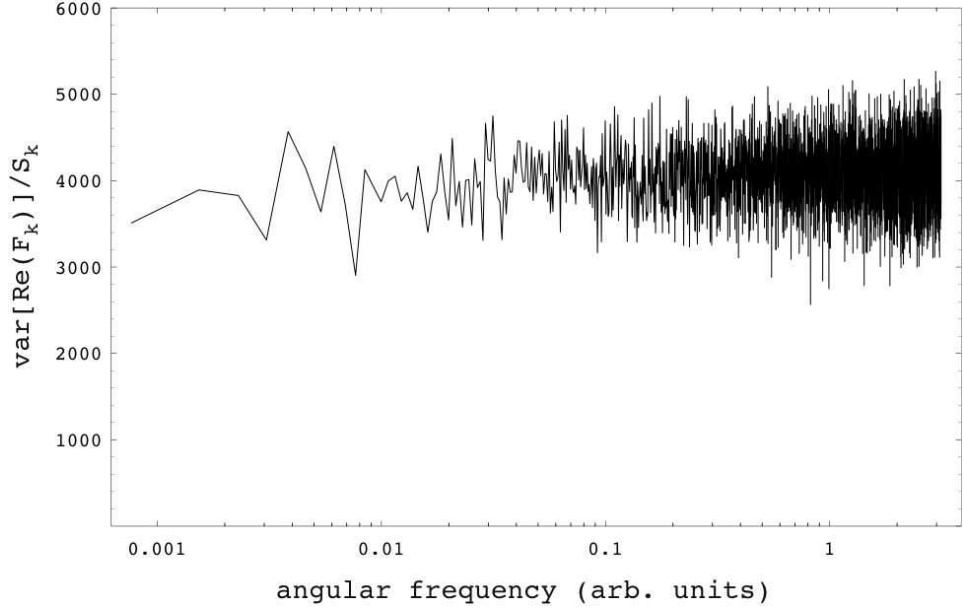}
\caption{\label{fig14} Ratio $\mathrm{var} [\Re (F_k)] /S_k$ obtained from $2^{20} = 1048576$ samples with $A=1$ and a uniform distribution of decay rates, in the range  $\lambda_{min} = 0.0001$, $\lambda_{max} = 1$, averaged over the DFT results obtained from 128 blocks of 8192 samples each. The average ratio fluctuates about constant level, and this is in line with the usual hypothesis that $\mathrm{var} [\Re (F_k)] \propto S_k$ (see, e.g., \cite{TK})).
}
\end{figure}

This generator can be used to test a standard hypothesis that is commonly used with FFT-based colored noise generators, in analogy to the well-known behavior of white noise, namely that the standard deviation of the real and imaginary parts of the discrete Fourier components $F_k$ of a colored noise process is proportional to the square root of the noise spectrum $S_k$  \cite{TK}. Figure \ref{fig14} shows the ratio $\mathrm{var} [\Re (F_k)] / S_k$ for a simulated $1/f$ noise: the average ratio is constant and thus the simulation does not disprove the standard assumption, at least for this particular noise process.

In all the examples described above the sampling interval $\Delta s$ is fixed, but the method is in no way limited to constant sampling intervals. And indeed this is probably the greatest strength of this noise generator, its ability to work also with uneven sampling intervals: this is not true for the other common generators \cite{kas}.

To conclude, in this paper I have described a generator of colored noise that is exact, is not limited to evenly distributed samples, has a well-behaved complexity $O(N_s)$ (in contrast to many other generators that have a $O(N_s \log N_s)$ complexity), and is not troubled by hidden periodicity issues, like the FFT-based generators.

\pagebreak


\end{document}